\documentclass[5p,twocolumn,times,number]{elsarticle}

\usepackage{graphicx}
\usepackage{amsmath}   
\usepackage{lineno}

\begin{document}

\begin{frontmatter}

\title{The Barrel DIRC detector of PANDA }
\author[a]{C.~Schwarz\corref{cor}} 
\ead{C.Schwarz@gsi.de}
\author[a,b]{A.~Ali}
\author[a]{A.~Belias} 
\author[a]{R.~Dzhygadlo,} 
\author[a]{A.~Gerhardt} 
\author[a,b]{M.~Krebs}
\author[a]{D.~Lehmann} 
\author[a,b]{K.~Peters}
\author[a]{G.~Schepers} 
\author[a]{J.~Schwiening} 
\author[a]{M.~Traxler}
\author[c]{L.~Schmitt}
\author[d]{M.~B\"{o}hm}
\author[d]{A.~Lehmann}
\author[d]{M.~Pfaffinger}
\author[d]{F.~Uhlig}
\author[d]{S.~Stelter}
\author[e]{M.~D\"{u}ren}
\author[e]{E.~Etzelm\"{u}ller}
\author[e]{K.~F\"{o}hl}
\author[e]{A.~Hayrapetyan}
\author[e]{K.~Kreutzfeld}
\author[e]{J.~Rieke}
\author[e]{M.~Schmidt}
\author[e]{T.~Wasem}
\author[f]{P.~Achenbach}
\author[f]{M.~Cardinali}
\author[f]{M.~Hoek}
\author[f]{W.~Lauth}
\author[f]{S.~Schlimme}
\author[f]{C.~Sfienti}
\author[f]{M.~Thiel}


\cortext[cor]{Corresponding author}

\address[a]{GSI Helmholtzzentrum f\"ur Schwerionenforschung GmbH, Darmstadt, Germany}
\address[b]{Goethe University, Frankfurt a.M., Germany}
\address[c]{FAIR, Facility for Antiproton and Ion Research in Europe, Darmstadt, Germany}
\address[d]{Friedrich Alexander-University of Erlangen-Nuremberg, Erlangen, Germany}
\address[e]{II. Physikalisches Institut, Justus Liebig-University of Giessen, Giessen, Germany}
\address[f]{Institut f\"{u}r Kernphysik, Johannes Gutenberg-University of Mainz, Mainz, Germany}

\begin{abstract}

The PANDA experiment is one of the four large experiments being built at
FAIR in Darmstadt. It will use  a cooled antiproton beam on a fixed target within
the momentum range of 1.5 to 15 GeV/c to address questions of strong QCD,
where the coupling constant $\alpha_s \gtrsim 0.3$.
The luminosity of up to $2 \cdot 10^{32} cm^{-2}s^{-1}$ and the momentum resolution of
the antiproton beam down to
\mbox{$\Delta$p/p = 4$\cdot10^{-5}$} allows for high precision spectroscopy, especially for rare 
reaction processes.
Above the production threshold for
open charm mesons the production of kaons plays an important role for identifying the
reaction. The DIRC principle allows for a compact
particle identification for charged particles in a hermetic detector,
limited in size by the electromagnetic lead tungstate calorimeter. The Barrel DIRC
in the target spectrometer covers polar angles between $22^\circ$ and
$140^\circ$ and will achieve a pion-kaon separation of 3 standard deviations
up to 3.5 GeV/$c$. Here, results of a test beam are shown for a single radiator bar
coupled to a prism with $33^\circ$ opening angle, both made from synthetic
fused silica read out with a photon detector array with 768 pixels.

\end{abstract}


\end{frontmatter}

\section{Introduction}

The PANDA experiment~\cite{panda1} employs several particle identification (PID) systems for its physics program
\cite{panda-physics}. Two of them are 
Cherenkov counters using the DIRC (Detection of Internally Reflected Cherenkov light)
principle which allow for building compact PID systems. This is
accomplished by using the radiators as light guides for the produced Cherenkov photons
to a photon detector outside of the active volume. The first of such detectors was built
and successfully operated by the BaBar experiment~\cite{coyle2004,adam2005}.
The Barrel DIRC described here, separates pions from kaons with at least 3 standard
deviations (s.d.) up to momenta of \mbox{$3.5$~GeV/$c$}.

\section{Design}
The design of the PANDA Barrel DIRC  \cite{schwiening2018,DIRC_TDR} is based
on the design of the BABAR DIRC with several
improvements. The radius of the 16-sided polygonal barrel is 476~mm.
\begin{figure}
\centering
\includegraphics[width=0.80\linewidth]{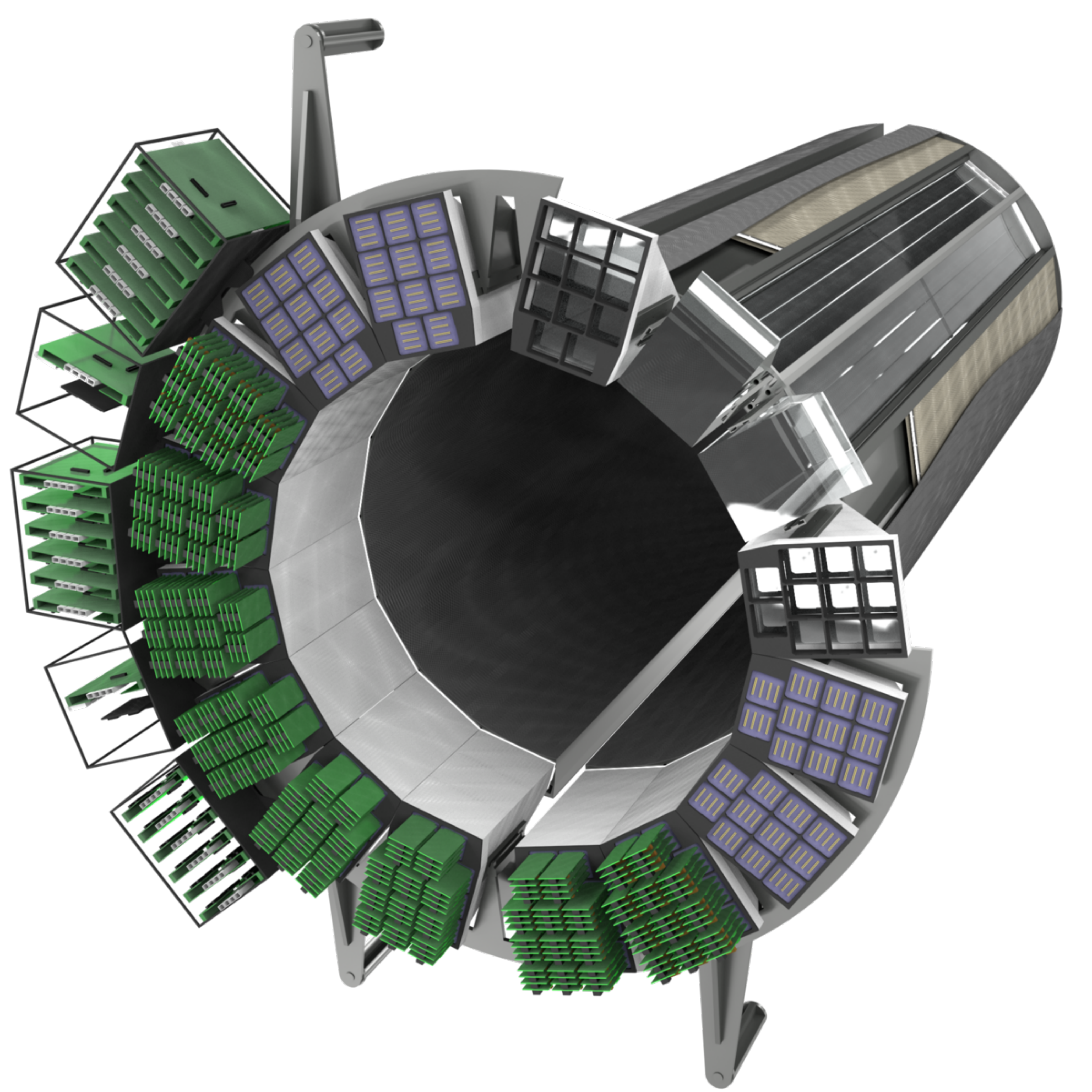}
\caption{Schematic of the Barrel DIRC baseline design \cite{schwiening2018}.}
\label{fig:barrel_dirc}
\end{figure}
To eliminate photon background from neutrons hitting the large water
tank between the 
radiators and photon detectors observed by BaBar 
it was replaced by 16 smaller $30$~cm-deep  prisms made from
synthetic fused silica. Coupled to each prism are three radiator bars and
a photon detector array of eleven lifetime-enhanced Microchannel-Plate (MCP) PMTs
\cite{lehmann2016} on the opposite readout side.
Each radiator bar has a thickness  17~mm, a width of 53~mm, 
and a length of 2400~mm.
The DIRC detector is partitioned in 16 independent segments with
radiators  and an expansion volume with the photon detector that can be dismounted from the radiators. 
The separation
into smaller "cameras" and the usage of synthetic fused silica was already favored by 
the SuperB FDIRC \cite{dey2015}  and the Belle~II TOP \cite{krizan2014}. The reduction
in size and the finite radiator bar size requires focusing the photons
onto the photon detector. Space limitations within the magnetic yoke of the target
spectrometer solenoid excludes the usage of focusing mirrors.   A three-layer 
spherical compound lens between radiator bar and expansion prism is used to get a sharp
Cherenkov image. The readout electronics is based on the
HADES Trigger Readout Board (TRB)~\cite{trb3-jinst}
and front end amplification and discriminator cards directly mounted on the
MCP-PMTs\cite{cardinali:padiwa}.

\section{Beam tests}
\begin{figure}
\centering
\includegraphics[width=0.95\linewidth]{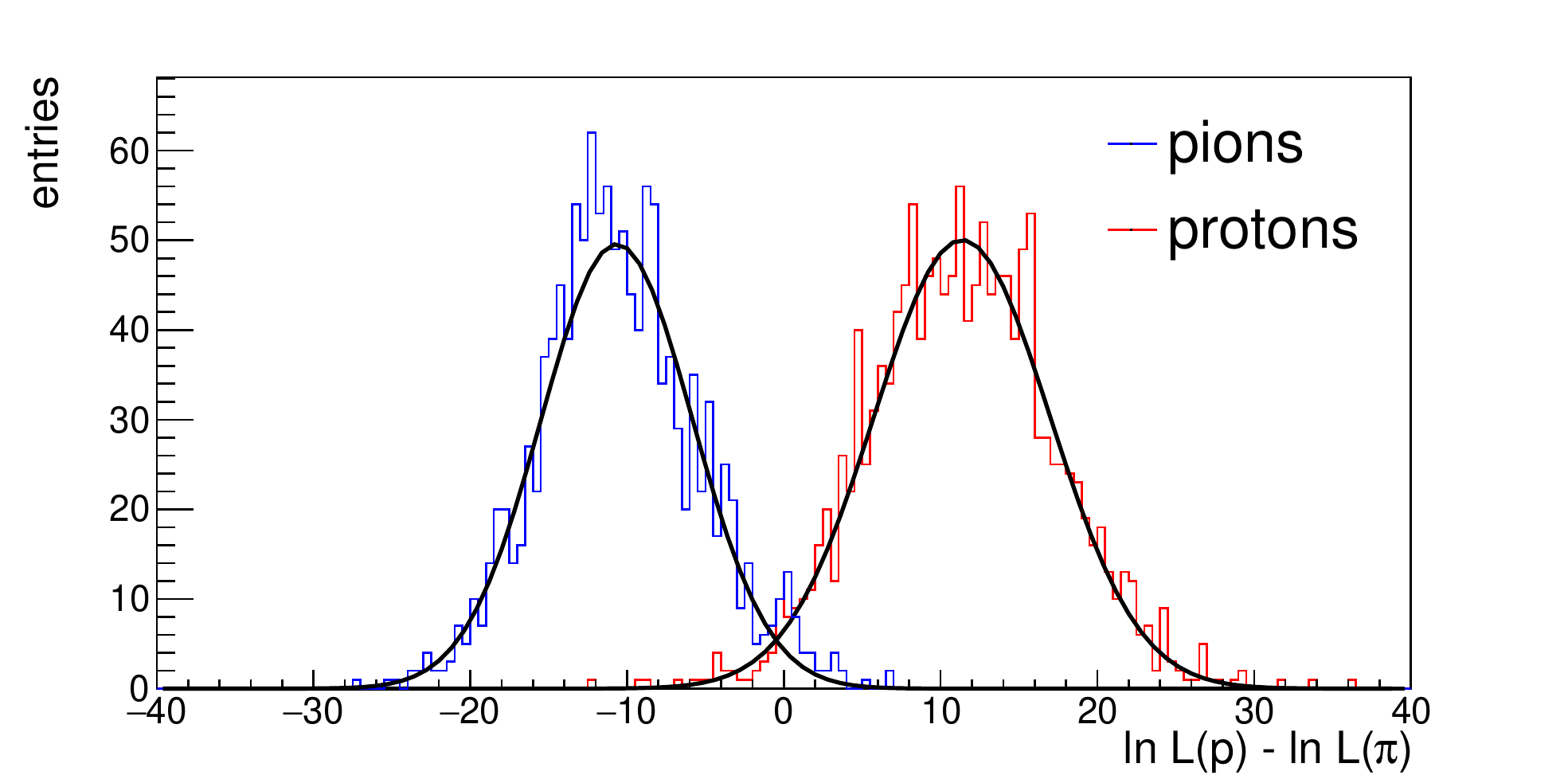}
\caption{
One of the figure of merits for the performance of the Barrel DIRC is the 
proton-pion log-likelihood difference distributions for proton-tagged (red) and
pion-tagged (blue) beam events. Here, the result for the focusing with the three-layer spherical lens
is shown.}
\label{fig:sd}
\end{figure}

After several test beam campaigns at GSI and CERN in the recent years,  
the campaign at 
the CERN PS 2017 served the  validation of the baseline design together with slight 
changes of the setup
compared to previous beam times. 
The prototype consisted of the important elements for one bar box: 
A fused silica bar 
(17.1 $\times$ 35.9 $\times$ 1200.0~mm$^3$) was coupled on one end to a flat mirror, 
on the opposite end to a focusing lens and the fused silica prism as optical expansion volume.
Latter had an opening angle of 33$^\circ{}$  and allowed, different from
previous beam times \cite{schwiening2018,schwarz2017}, for coupling 12 MCP-PMTs  
with 64 pixels each. 
Two focusing lenses, a spherical and a cylindrical one,
made from LaK33B \cite{Schott} glass were used for tests.
Proton and pion beam particles were tagged by a time-of-flight system.
The reconstructed particle type comes from the comparison of the arrival time of the photons
at each MCP-PMT pixel and the expected arrival time from simulation (time-based imaging).
The log-likelihood difference distribution obtained using time-based imaging for a 
polar angle of 25$^\circ$ and a momentum of 7 GeV/c is shown in Fig. \ref{fig:sd} 
together with Gaussian fits. 
A $\pi/p$ performance of 4.0 $\pm$ 0.1 s.d. is observed. It 
deteriorates to 3.0 $\pm$ 0.1 s.d. when using a three-layer cylindrical lens for focusing.
Simulation \cite{DIRC_TDR,geant4} are used to extrapolate the result for the three-layer spherical lens 
to the expected $\pi/K$ separation power. 
With the expected photon detection efficiency and timing precision of
the MCP-PMTs in the Barrel DIRC, a $\pi/K$ separation power of 
about 6.6 s.d. at 25$^\circ$ and 3.5 GeV/$c$ is achieved. This excellent result allows for studying the
performance of the setup with a reduced number or photon detectors in coming 
experiments to achieve a cost optimization.

\section*{Acknowledgments}

This work was supported by 
HGS-HIRe, 
HIC 
for FAIR, 
BNL eRD14.
We thank the GSI and CERN staff for the opportunity to use 
the beam facilities and for their on-site support.


\end{document}